\documentclass[fleqn,10pt]{wlscirep}
\graphicspath{{./}{SR_figures_with_landscape/}}
\usepackage[utf8]{inputenc}
\usepackage[T1]{fontenc}
\title{Controlling Inter-Particle Distances in Crowds of Motile, Cognitive, Active Particles}

\author[1]{Rajendra Singh Negi}
\author[1]{Priyanka Iyer}
\author[1,*]{Gerhard Gompper}
\affil[1]{Theoretical Physics of Living Matter, Institute of Biological Information Processing and Institute of Advanced Simulation, Forschungszentrum J{\"u}lich, 52425 J{\"u}lich, Germany}

\affil[*]{g.gompper@fz-juelich.de}

\begin{abstract}
Distance control in many-particle systems is a fundamental problem in nature.
This becomes particularly relevant in systems of active agents, which can sense
their environment and react by adjusting their direction of motion. 
We employ agent-based simulations to investigate the complex interplay 
between agent activity, characterized by P{\'e}clet number $Pe$, reorientation maneuverability $\Omega$, 
vision angle $\theta$ and vision range $R_0$, and agent density, which determines agent 
distancing and dynamics. We focus on semi-dense crowds, where the vision range is much larger than
the particle size. The minimal distance to the nearest neighbors, exposure time, and persistence of orientation 
direction are analyzed to characterize the behavior. With increasing particle speed at fixed maneuverability, 
particles approach each other more closely, and exhibit shorter exposure times. The temporal persistence of 
motion decreases with increasing $Pe$, reflecting the impact of activity and maneuverability on direction 
changes. For a vision angle $\theta=\pi/4$, we observe the emergence of flocking aggregates with a 
band-like structure, reminiscent of the Viscek model. Additionally, for vision angles $\theta\ge \pi/2$, 
several quantities are found to display a universal scaling behavior with scaling variable $Pe^{3/2}/\Omega$. 
Our results are in good agreement with recent experiments of pedestrians in confined spaces.
\end{abstract}
\begin{document}

\flushbottom
\maketitle
%
%
\thispagestyle{empty}

\section*{Introduction}

Controlling and keeping distances is a ubiquitous issue, both in condensed matter and in living systems. 
In liquid or crystalline condensed phases at thermal equilibrium, the distance between neighboring atoms 
or molecules is determined by the competition of short-range repulsive and longer-range
attractive interactions \cite{chai95}. 
In colloidal systems, interactions can be designed in many ways, and systems with unusual interactions, 
like short-range attractive and long-range repulsive, have been constructed \cite{scio04}. The well-controlled condensed 
phases are important for many bulk material properties, like compressibility, shear modulus, electrical 
conductivity, etc. Interestingly, also purely repulsive interactions can lead to crystallization, such 
as in a gas of electrons moving in a uniform, inert, neutralizing background, where optimal
distancing, determined by a minimum of the electrostatic energy, is found to be attained by the formation 
of a lattice structure -- the Wigner crystal -- if the electron density is less than a critical threshold
\cite{wign34}.  Similarly, the maximization of distance under some constraints, 
such as in the Thomson problem of the distribution of electrons on the surface of
a sphere, can lead to crystallization with topological defects \cite{bowi02}. 

The problem of controlling and optimizing distance becomes much more complex and interesting in motile 
active and living systems \cite{elgeti_2015_RPP}. A simple -- one-dimensional -- example is traffic flow on a highway.
Here, distances between cars have to exceed the minimal breaking distance, which grows with 
increasing speed $v_0$, quadratically for the stopping distance, linearly in flowing traffic \cite{coif15}. 
This implies an optimal distance to maximize flow, the product of speed and density \cite{helb09}. 
In many living systems, where motion typically occurs in two or three spatial dimensions, 
distances between individuals should not be too large to facilitate mating and reproduction, and 
to collectively protect a group against predators \cite{diabate2011spatial,foster1981evidence,olson2013predator}. At the same time, distances should not 
be too small so as not to hinder the search for food, or the individual motion, or even damaging 
collisions. Also, to prevent the spread of airborne infectious diseases, like COVID-19, it is important 
to maximize the distance to other individuals and to avoid crowded spaces 
\cite{ciotti2020covid, pouw2020monitoring, chraibi_2023_social}. However, recent studies of a model of  
active motion of finite-size particles with constant speed $v_0$ and slow rotational 
diffusion -- called active Brownian particles (ABPs) -- shows that activity can have the 
the opposite effect of  motility-induced clustering and phase separation \cite{cate15}. 
The origin of this behavior is the formation of small clusters by head-on collisions of a few particles, 
which only slowly disintegrate and thereby form the nucleus of larger clusters.

The essential difference between APBs and living individuals, such as birds or pedestrians, is, 
of course, that the former are ``dumb", while the latter have a visual perception of their 
environment, and use this information to react by adapting their speed and direction of motion to 
avoid collisions.
A pivotal issue revolves around the efficacy of individual pedestrians in upholding interpersonal 
distancing within dense crowds \cite{schadschneider2008evacuation, chraibi2010generalized, lu2021_jsm}. 
Recent controlled laboratory experiments with pedestrians moving in a room 
\cite{Echeverria_2021_SR, echeverria_2022_PRE} have cast light on the implications of factors 
such as pedestrian density, walking speed, and prescribed safety distances on interpersonal spacing 
within moderately crowded environments. 

In this study, we aim to elucidate the physical mechanisms underlying the cognitive self-steering of
pedestrians (or birds) in moderately dense crowds with nearly homogeneous spatial distribution. 
We consider a highly simplified model of cognitive self-steering particles (intelligent active Brownian 
particles, iABPs), which move with constant speed $v_0$, can sense their environment by visual perception, 
and react by applying a limit steering torque (``maneuverability"),
but have no memory (and thus cannot estimate the speed of neighboring  particles nor their direction 
of motion) \cite{barberis_2016_PRL, goh_2022_NJP, negi_2022_soft_matter}. Thus, our iABPs have to 
base their decisions on the redirection of motion entirely on the the instantaneous position of 
their neighbors. Similar self-steering mechanisms have been considered in models of social 
interactions in animal groups \cite{couzin_2005_nature}. We want to emphasize that our torque-based 
steering mechanism is different from the short-distance repulsion of some other swarming and flocking 
models, which employ a conservative repulsive interaction potential \cite{d2006self}.
We perform extensive agent-based simulations to analyze iABP dynamics at finite density, 
in order to explore the complex interplay between particle density, activity level, maneuverability, 
vision angle, and vision range. Key factors such as the distance to the nearest neighbors, exposure time, 
and persistence in velocity direction are analyzed. The simulation results are compared with
the results of recent experiments on pedestrians in a room, to gain insights to which extent our simple 
model is able to reproduce and explain pedestrian behavior under the imperative of maximizing distance.

\section*{Results}
\subsection{Model and Simulation Approach}
\label{sec:model}
We consider a system of $N$ agents which are modeled as point particles.
The equation of motion of particle $i$ with position $\boldsymbol{r}_{i}$ is 
\begin{equation} \label{pos_equation}
    m \ddot{\mathbf{r}}_i =- \gamma \dot{\mathbf{r}}_i + \mathbf{F}_{\text{act}} \mathbf{e}_i.
\end{equation}
Here, $m$ is the mass of the particle, $\gamma$ the translational friction coefficient, and 
$\mathbf{F}_{\text{act}}$ the propulsion force along the instantaneous particle
orientation $\mathbf{e}_i$, resulting in the overdamped limit in a constant velocity 
$v_0=|F_{act}|/\gamma$. The self-steering behavior of each agent is affected 
by the positions of neighboring particles. Particle $i$ can adjust its propulsion direction 
$\mathbf{e}_i$ through self-steering in the direction 
$\mathbf{u}_{ij} = (\mathbf{r}_j - \mathbf{r}_i)/|\mathbf{r}_j - \mathbf{r}_i|$, determined by its 
neighbors, with an adaptive torque $\mathbf{M}^{av}_i$, as 
\cite{negi_2022_soft_matter, negi2023collective, barberis_2016_PRL}
\begin{equation}
    \dot{\mathbf{e}}_i(t) = \mathbf{M}^{av}_i + \boldsymbol{\Lambda}_i(t) \times \mathbf{e}_i(t),
\end{equation}
where $\boldsymbol{\Lambda}_i$ represents Gaussian and Markovian stochastic processes with zero mean and 
correlations $\langle \boldsymbol\varLambda_i(t) \cdot \boldsymbol\varLambda_j (t') \rangle = 2(d-1)D_R \delta_{ij} \delta(t-t')$ 
in $d$ spatial dimensions with rotational diffusion coefficient $D_R$.

\begin{figure*}
    \centering
    \includegraphics[width=.85\textwidth]{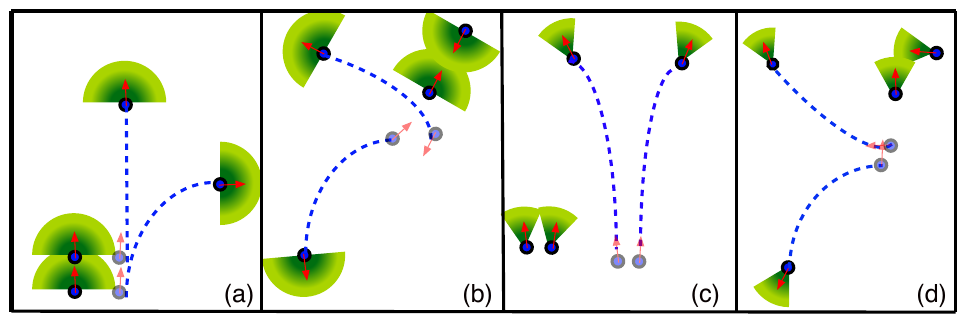}
    \caption{Schematic representation of two-particles interaction through visual perception
    and self-steering to avoid a close approach, as described by Eq.~\eqref{eq:steering}. The field of vision 
    is colored green, and corresponding trajectories are indicated by dashed lines. The initial configuration
    of the vision cones are illustrated near the starting points (with the initial propulsion direction indicated 
    by semi-opaque arrows) of the trajectories, while the final configurations  accompany the trajectory lines.}
    \label{fig:steering}
\end{figure*}

The cognitive torque (referred to as the "visual" torque) acting on particle  $i$ is
\begin{equation}
\label{eq:steering}
     \mathbf{M}^{av}_i = -\frac{C_0}{N_{c,i}} \sum_{j\in VC} 
                      e^{-r_{ij}/R_0} \mathbf{e}_i \times (\mathbf{u}_{ij} \times \mathbf{e}_i),
\end{equation}
where $C_0$ represents the ``visual" maneuverability, and $N_{c,i}$ is the number 
of particles within the vision cone (VC). Particles within the VC are determined based on the 
condition $\mathbf{u}_{ij} \cdot \mathbf{e}_i \geq \cos(\theta)$, where $\theta$ is the vision angle, 
defining the opening angle of the vision cone centered on the particles's orientation vector $\mathbf{e}_i$. 
In addition, we limit the vision to $|\boldsymbol r_i - \boldsymbol r_j| \leq R_V$, where $R_V > R_0$ is the vision range. 
and treat all further apart particles as invisible, which is helpful for computational efficiency. 
Finally, the number of effectively visible particles in Eq.~\eqref{eq:steering} is
\begin{equation} 
\label{eq:eq_2_norm}
  N_{c,i} =  \sum_{j\in VC}e^{-r_{ij}/R_0} .
\end{equation}
Here, some comments are in order. First, it is important to note that the steering torque, Eq.~\eqref{eq:steering},
is {\em non-additive}, due to the normalization by the visible particle number $N_{c,i}$. 
Second, the exponential range $R_0$ can be understood as a reduced vision range at higher local density of 
neighboring particles, for example due to a blocking of the view on distant neighbors by those close by.
The employ here the choice $R_V=4R_0$ if not stated otherwise. Finally, the steering torque in Eq.~\eqref{eq:steering} 
implies an effective repulsive interaction, as illustrated schematically in Fig.~\ref{fig:steering}.

In polar coordinates in two spatial dimensions, $\mathbf{e}_i = (\cos \varphi_i , \sin \varphi_i)^T$, the 
equations of motion for the orientation angles $\varphi_i$ become 
\begin{equation}\label{eq:polar_eq_fv}
     \dot{\varphi}_i = -\frac{C_0}{N_{c,i}} \sum_{j\in VC} e^{-r_{ij}/R_0} \sin({\phi_{ij}-\varphi_i}) + \Lambda_i(t).
\end{equation}
The sum on the right-hand side of Eq.~\eqref{eq:polar_eq_fv} describes the tendency of a particles to move 
away from regions of high local particle density within its vision cone (VC).

In simulation, we measure time in units of $\tau_R$, length in units of $R_0$, but keep these units 
explicit in all expressions. The activity of the pedestrians is given by the P{\'e}clet number
\begin{equation}
    Pe = \frac{v_{0}}{R_{0} D_{R}}, 
\end{equation}
while the scaled maneuverability is  
\begin{equation}
    \Omega= C_0/D_{R}.
\end{equation}
Periodic boundary conditions of a square simulation box of linear extension $L$ are employed to control 
the dimensionless particle density $\Phi=N (R_0/L)^2$. Compared to a system with explicit walls -- 
unavoidable in experiments, like those with pedestrians \cite{Echeverria_2021_SR} -- this has the 
advantages that the system is completely homogeneous, and that particle motion over long distances 
can be analyzed. 

We want to emphasize that (i) all particles move with constant velocity, no 
speed adaptation is considered, and (ii) no volume exclusion of particles is taken into account, 
in order to avoid jamming, which corresponds to systems for which the vision range is much larger 
than the particle size.  Thus, our model applies to semi-dense crowds.

The simulations are performed in the over-damped limit, i.e. $m D_{R} /\gamma\ll 1$ 
, so that inertial effects are negligible.
Explicitly, we choose $\gamma= 10^{2} D_{R} $ and $m=1$. The linear dimension of the simulation box 
is $L/R_{0} = 20$. We study systems with particle numbers 
$N=25$, $64$, $100$, and $225$, which corresponds to densities 
$\Phi=0.0625$, $0.16$, $0.25$, and $0.5625$, respectively. 
The equations of motion \eqref{pos_equation} are solved with a  velocity-Verlet-type 
algorithm suitable for stochastic systems \cite{gronbech_2013_MP},  with the time step 
$\Delta t = 10^{-3} / D_{R}$.  

\subsection{Distance to Nearest Neighbors}
\label{sec:ped:Distance_NND1}
We analyze the probability distribution functions (PDFs) for the distance $d_1$ to 
the nearest neighbor,  and extract information on the average 
$\langle d_1 \rangle$ and the fraction of particles closer than $R_0$ to other 
particles. 
We focus on the dependence on key parameters, like particle density $\Phi$, P{\'e}clet 
number $Pe=v_0/(R_0 D_R)$, maneuverability $\Omega$, and vision angle $\theta$.
\subsubsection{Effect of Particle Density, Activity, and Maneuverability} 
\label{sec:d1_NPe}
\begin{figure}
    \centering
     \includegraphics[width= \textwidth]{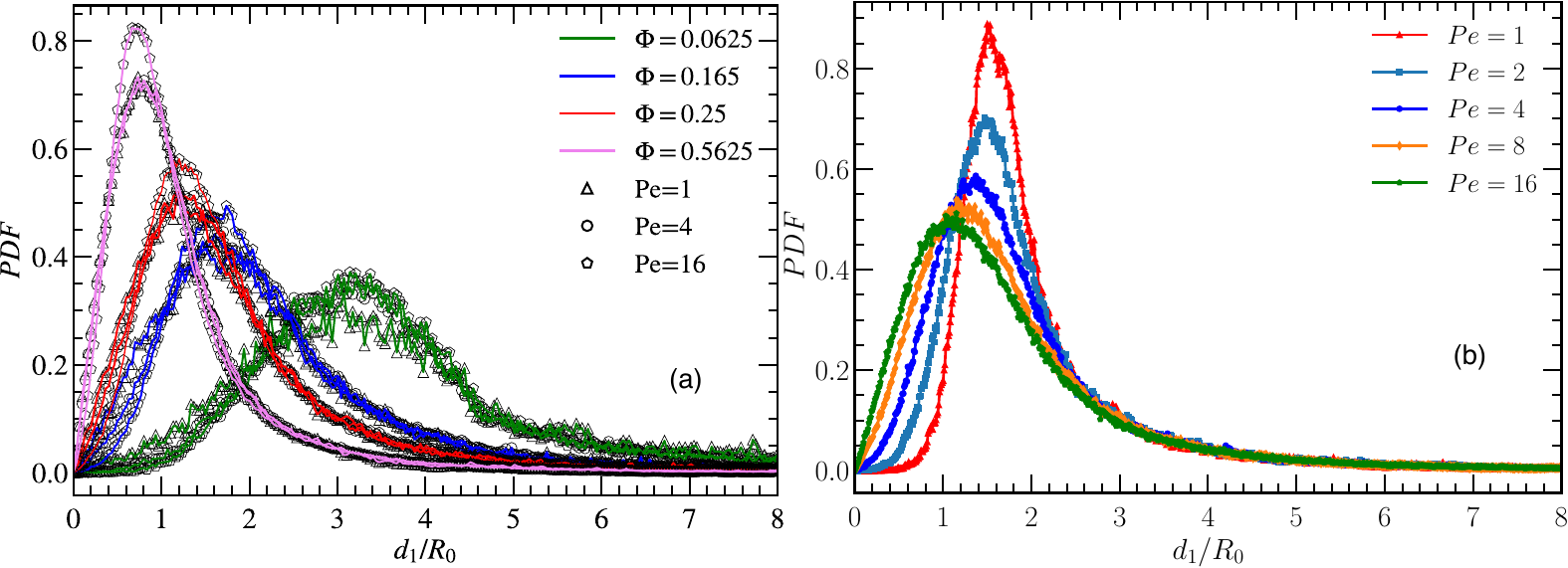}
    \caption{Probability density function (PDF) of the distance $d_1$ to the nearest neighbor. 
    (a) For various particle densities $\Phi = 0.0625, 0.165, 0.25, 0.5625$ and activities   
    $Pe \in{(1,4,16)}$, with fixed
    activity-to-maneuverability ratio $Pe^{3/2}/\Omega=1$, and vision angle $\theta=\pi/2$. 
    (b) Dependence on P{\'e}clet number $Pe$, with fixed particle density $\Phi=0.25$, maneuverability 
    $\Omega=16$, vision angle $\theta=\pi/2$, cutoff range $R_V= 4 R_{0}$.  
    .}
    \label{fig:PDF_PSD}
\end{figure}
Figure~\ref{fig:PDF_PSD}(a) shows the probability density functions (PDFs) of 
nearest-neighbor distance $d_1$ for various particle densities
and activities $Pe$, for fixed vision angle $\theta=\pi/2$. The increase in 
the particle density from $\Phi=0.0625$ to $\Phi=0.5625$ results in a shift of the 
distribution towards lower distances $d_1$, indicating closer approaches between 
particles in more crowded environments. This results from a reduction in the inter-particle distance with density as $\Phi^{-1/2}$ -- independent of particle mobility. Another interesting result is that a constant ratio $Pe^{3/2}/\Omega$ (in this case $Pe^{3/2}/\Omega=1$) results in a collapse of the distributions onto a single master
curve, which indicates a tight coupling of individual activity and maneuverability. 
This is due to the requirement of higher steering torques for larger particle 
speed;  a similar behavior has been found previously for pursuit dynamics, where scaling 
with $Pe/\Omega^{1/2}$ is observed \cite{goh_2022_NJP, gassner_2023_epl}. 
The $Pe^{3/2}/\Omega$ scaling will be discussed in more detail in the context of the
average minimal distance $\langle d_1 \rangle$ below. 

To study the effect of activity $Pe$, we analyze distance distribution $P(d_{1})$  at fixed $\Omega$ 
and particle density, see Fig.~\ref{fig:PDF_PSD}(b) for vision angle $\theta=\pi/2$, at various activities. 
Particles come closer to each other for larger $Pe$.
Thus, slower-moving particles can maintain a larger distance because they can steer away from other 
particles already at a larger distance -- at constant maneuverability.   

Similar behavior is reported for pedestrians, where slower-moving can maintain a higher distance amongst 
themselves in comparison with faster-moving pedestrians \cite{Echeverria_2021_SR}. 

\begin{figure}
    \centering
    \includegraphics[width=.95\textwidth]{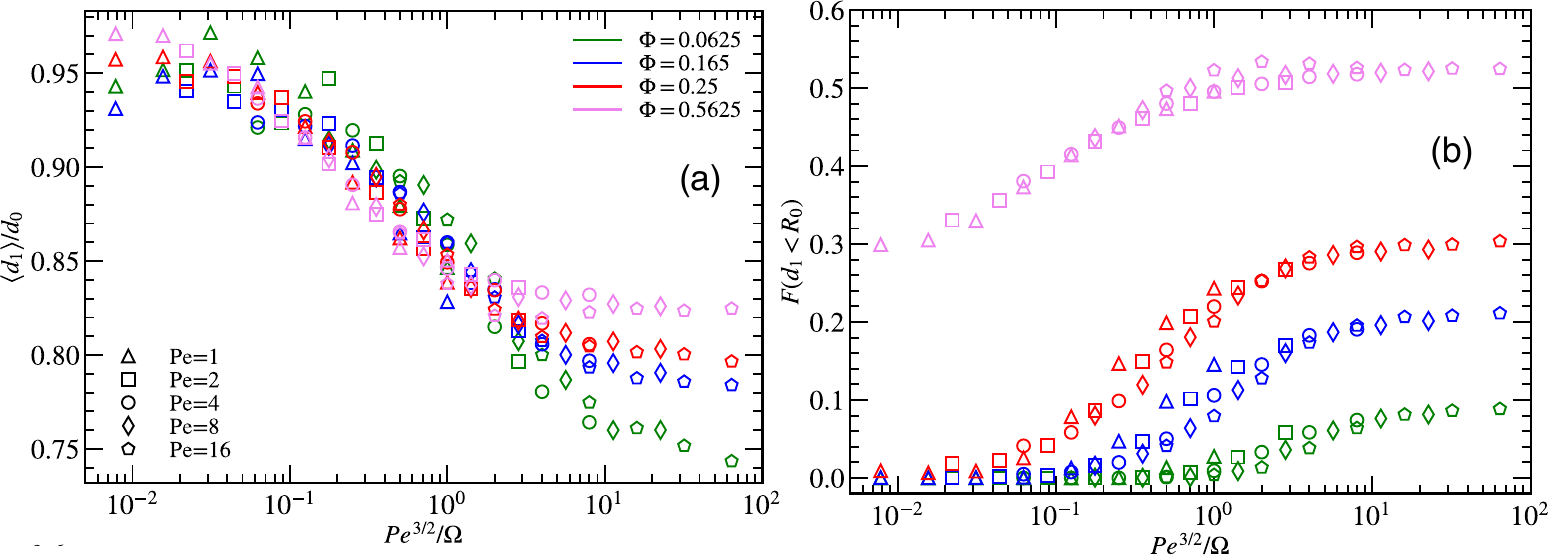}
    \caption{
    (a) Average nearest-neighbor distance $d_1$ for various pedestrian densities $\Phi$ 
    as a function of $Pe^{3/2}/\Omega$.  
    (b) Fraction of particles within a distance $R_{0}$ from other particles as a 
    function of $Pe^\beta/\Omega$, with $\beta=3/2$, for various 
    particle number $N$, as indicated. The vision angle in both cases is $\theta=\pi/2$. 
    }
    \label{fig:Mean_D1_F(d1)_N_Pe_varation}
\end{figure}

From the PDF $P(d_{1})$, we can calculate 
the average minimal distance $\langle d_{1} \rangle$ to nearest neighbors, and the fraction 
of particles, $F(d_{1}<R_{0})$, which are at a distance to their nearest neighbors 
less than $R_{0}$. 
Figure \ref{fig:Mean_D1_F(d1)_N_Pe_varation}(a) shows  $\langle d_1 \rangle$, scaled with 
the neighbor distance $d_{0}= 2L/ \sqrt{\pi N'}$ in a regular triangular lattice with the same particle
density.  Here, we employ the effective particle number $N'=N+N_0$, with $N_0=3$ to account for 
finite-size effects and to improve the scaling. 
Figure \ref{fig:Mean_D1_F(d1)_N_Pe_varation}(a) demonstrates that the data for 
$\langle d_1 \rangle/d_{0}$ as a function $Pe^{3/2}/\Omega$ collapse reasonably well onto 
a universal scaling curve, as expected from the scaling of $P(d_{1})$.
Thus, the minimum distance $d_1$ decreases as the particle density increases as 
$\langle d_{1} \rangle \sim 1/\sqrt{N} \sim \sqrt{\Phi}$.  

Furthermore, the results of Fig.~\ref{fig:Mean_D1_F(d1)_N_Pe_varation}(a) indicate that three
dynamic regimes can be distinguished: 
\begin{itemize}
\item The ``overcautious distancing" regime [see movie M1 \cite{sup_mater}], for $Pe^{3/2}/\Omega \lesssim 0.05$, where particles 
keep a nearly constant distance from all neighbors, at the cost of hardly any 
translational motion,
\item the ``wiggling and squirming'' regime [see movie M2], for $0.05 \lesssim Pe^{3/2}/\Omega \lesssim 10$, 
where steering helps particles to avoid each other while allowing significant 
persistent motion, and
\item the ``reckless motion" regime, for $Pe^{3/2}/\Omega \gtrsim 10$, where 
particles move without taking much -- or any -- notice of their neighbors.
\end{itemize}

In the ``wiggling and squirming" regime, particles tend to approach each other closely 
before initiating 
steering maneuvers to avoid collisions, consequently leading to a reduction in the 
distance between the closest neighbors with increasing $Pe^{3/2}/\Omega$. The plateau of $\langle d_{1} \rangle/d_0$ observed in the "reckless motion" regime at $Pe^{3/2}/\Omega \gtrsim 10$, aligns with the measured values obtained from the 'non-interacting' ABP simulations. This shows that for low maneuverability or high activity, agents do not react to each other and have limited scope to modify their movement direction.

Figure \ref{fig:Mean_D1_F(d1)_N_Pe_varation}(b) displays the fraction $F$
of particles, which have a distance less than $R_{0}$ to other particles. This fraction 
is examined for various particle densities and P{\'e}clet numbers $Pe$ for fixed maneuverability $\Omega$. 
The data for different $Pe$ also collapse onto a single master curve when plotted as a function of 
the scaling variable $Pe^{3/2}/\Omega$. The fraction $F$ of close neighbors attains its maximum/minimum 
when the particle density is high/low -- as to be expected because all distances decrease with
increasing particle density. 
Furthermore, $F$ is a monotonically increasing function of $Pe^{3/2}/\Omega$, consistent with 
behavior of $P(d_1)$ and $\langle d_{1} \rangle$.  
For large $Pe^{3/2}/\Omega$, the fraction $F$ gradually approaches a plateau. The plateau values are 
approximately $F \simeq 0.1$ for $\Phi=0.0625$, $0.2$ for $\Phi=0.165$, $0.3$ for $\Phi=0.25$, and 
$0.55$ for $\Phi=0.5625$. 
This saturation behavior can be attributed to a balance between the density of 
pedestrians, their movement characteristics, and the chosen threshold distance 
$R_{0}$. Once the fraction of close encounters near these limits, the additional 
increase in $Pe^{3/2}/\Omega$ has a diminishing effect on the fraction $F$. 

 
%
\begin{figure}
    \centering
    \includegraphics[width=0.95\textwidth]{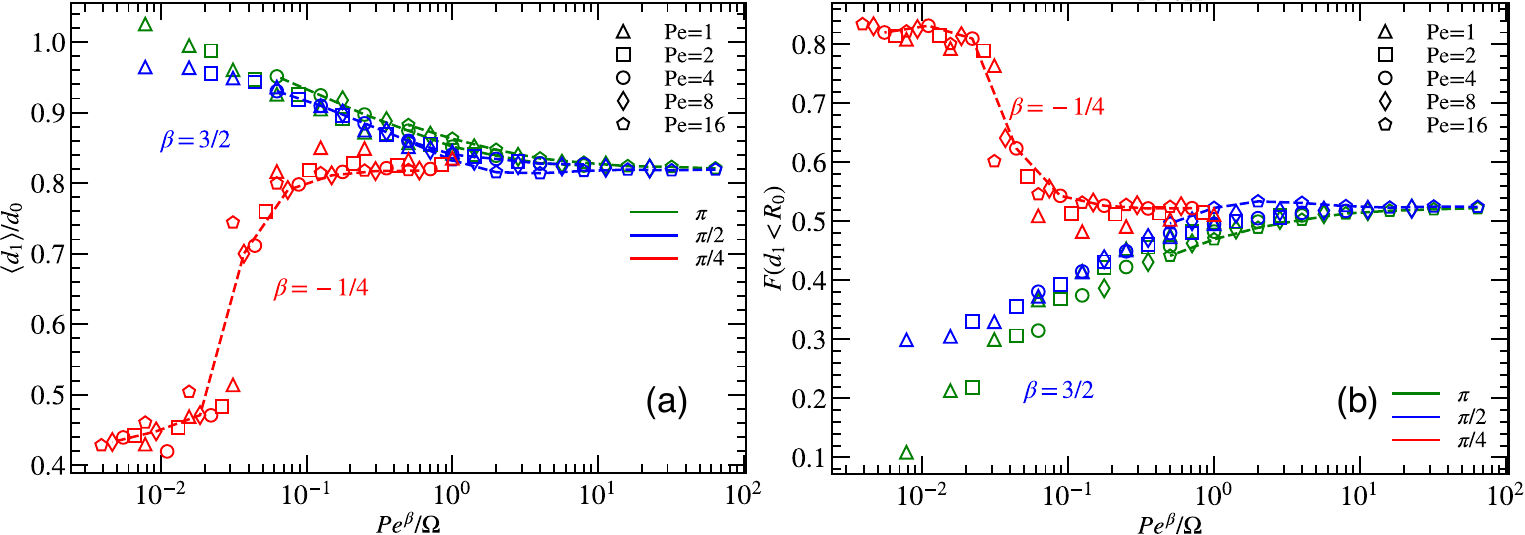}
    \caption{(a) Scaled average minimal distance $\langle d_1 \rangle$ for different vision angles $\theta$ 
    as function of 
    $Pe^{\beta}/\Omega$, where $\beta = 3/2$  for $\pi$ and $\pi/2$, and $\beta=-1/4$ for $\pi/4$. 
    (b) Fraction $F$ of particles within a distance $R_{0}$ from other particles  
    for various vision angle $\theta$ and the 
    $Pe^{\beta}/\Omega$ ratio. All data are for systems with particle density $\Phi=0.25$.}
    \label{fig:d1_angle}
\end{figure}

\subsubsection{Effect of Vision Angle}
\label{sec:d1_angle}

An important parameter of our model is the vision angle $\theta$.
Results for the average minimal distance $\langle d_1 \rangle$  between particles 
are displayed in Fig.~\ref{fig:d1_angle}(a) as a function of $Pe^{\beta}/\Omega$, where the
exponent $\beta$ is determined such as to optimize scaling with a single master curve. This yields 
$\beta \approx 3/2$ for vision angles $\pi$ and $\pi/2$, and $\beta \approx -1/4$ for vision angle $\pi/4$,
for fixed particle density $\Phi=0.25$ and vision range $R_{V}=4R_{0}$. The behavior for
large vision angle $\theta \ge \pi/2$ is found to be very different than for smaller vision angle $\theta=\pi/4$. 
For vision angle $\theta=\pi$, the behavior is essentially the same as for $\theta=\pi/2$ discussed
above, see Fig.~\ref{fig:Mean_D1_F(d1)_N_Pe_varation}. 
In particular, there is a good data collapse with scaling variable $Pe^{3/2}/ \Omega$. 

However, the situation changes quite dramatically for smaller vision angle $\theta=\pi/4$.
Here, the restricted field of view limits the particle's ability to detect each other from all directions. 
The restricted field of vision now does not always prevent collision, as particles can move toward each 
other with neither of them being able to the see the other. This can  
lead to very small particle  separation. The minimal distance  $\langle d_{1} \rangle / d_{0} $ now 
shows good data collapse with scaling variable $Pe^{-1/4}/\Omega$, see Fig.~\ref{fig:d1_angle}(a).
Note that since both $Pe$ and $\Omega$ contain a factor $1/D_R$, $Pe^{-1/4}/\Omega \sim D_R^{5/4}$,
so that the scaling variable depends strongly on the rotational diffusion.
This scaling with $Pe^{-1/4}/\Omega$ also implies that similar behavior is seen when $Pe$ and $\Omega$ 
are {\em inversely} proportional to each other, i.e. $Pe$ high, $\Omega$ low, and vice versa. 
The average minimal distance $\langle d_{1} \rangle/d_0$ attains a minimum for low $Pe^{-1/4}/\Omega$, which
corresponds to high values of both maneuverability and activity (and small $D_R$). This behavior arises from 
collective motion, where particles are moving in parallel as a band, very similarly as particles with 
alignment interactions in the Vicsek model. The reason is that for particles to stay in the band, they
need a high persistence of motion to retain their parallel motion, and a high maneuverability to be able to
quickly correct their direction of motion should their orientation deviate too much from parallelity.
The particles in band-like structures for vision angle $\pi/4$ are in 
closer proximity to each other, which is also seen in the close-neighbor fraction $F$, see 
Fig.~\ref{fig:d1_angle}(b). The fraction $F$ decreases with decreasing
$\Omega$ and $Pe$ (and increasing $D_R$). The Band-like structures are characterized in more detail 
in Sec.~\ref{Sec:Ped:Band} below.

\subsubsection{Effect of Vision Range}

\begin{figure}
    \centering
    \includegraphics[width=\textwidth]{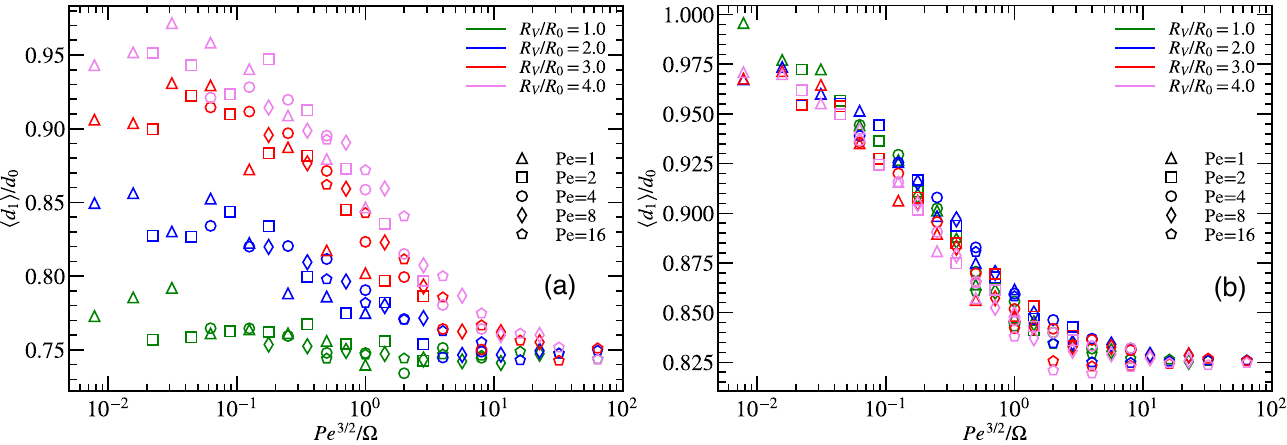}
    \caption{Average minimal distance $\langle d_1 \rangle$ for different $R_{V}$ and $Pe^{3/2}/\Omega$ for 
    particle density (a) $\Phi=0.0625$ and (b) $\Phi=0.5625$. Vision angle $\theta=\pi/2$ for all cases.}
    \label{fig:ped:vrange}
\end{figure}

In a model of visual-perception-induced steering, the vision range plays an essential role. Note that
some effect of the vision range has already been absorbed into the definition of the dimensionless
particle density $\Phi=N (R_0/L)^2$, where $R_0$ is the vision range at high local particle density.
Thus, to elucidate the effect of vision range, we now vary the ratio $R_V/R_0$.
Figure \ref{fig:ped:vrange} shows the average minimal distance $\langle d_1 \rangle$ for various $R_V/R_0$
ratios at low and high particle densities at fixed vision angle $\pi$. 
The qualitative behavior is similar for different vision ranges $R_V$. For low density $\Phi=0.0625$,
see Fig.~\ref{fig:ped:vrange}(a), $\langle d_1 \rangle$ for $Pe^{3/2}/\Omega \lesssim 1$ increases with 
increasing $R_V/R_0$, but saturates around $R_V/R_0 \simeq 4$. 
This happens because the effective density is now determined by $\Phi_V=N (R_V/L)^2$, with
$\Phi_V = 16 \Phi$ for $R_V/R_0=4$. Thus, the system is effectively at much higher density for large $R_V$. 

Conversely, in a dense system, see Fig.~\ref{fig:ped:vrange}(b), the effective vision range is essentially
represented by $R_0$, because the exponential factor in Eq.~\eqref{eq:steering} dominates, where particles 
beyond the distance $R_0$ hardly contribute. This leads a very weak dependence on the vision range $R_V$
already for $R_V/R_0 \gtrsim 1$.

\subsection{Exposure Time}
\label{sec:Exposure_T}

\begin{figure}
    \centering
    \includegraphics[width=0.45\textwidth]{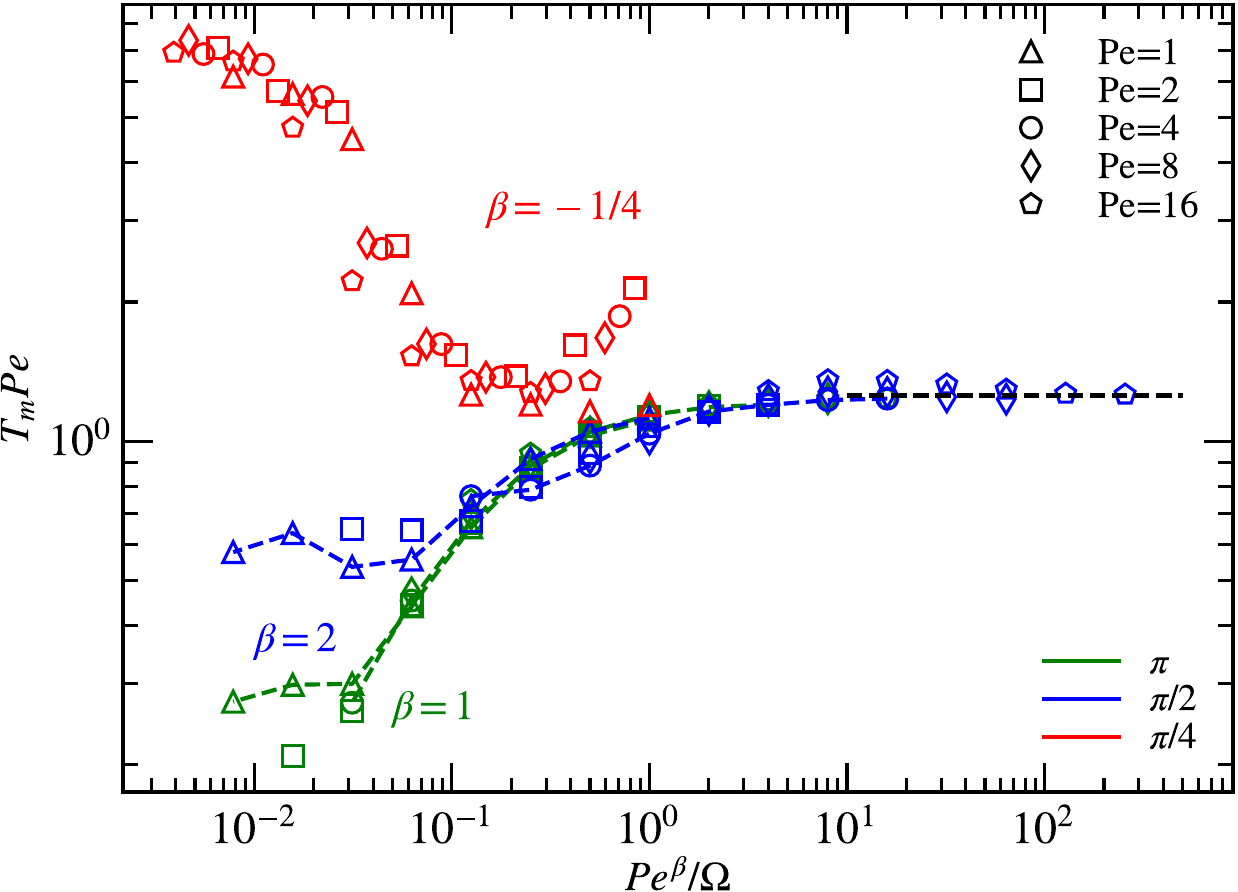}
    \caption{Scaled average exposure time, $T_m Pe$, during which particles remain close to each 
    other within a vicinity of radius $R_0$ uninterruptedly, for various P{\'e}clet 
    numbers ($Pe$), as indicated. 
     $\Phi=0.25$. }
    \label{fig:Expose_N_angle}
\end{figure}

Another interesting quantity to consider is the exposure time $T_{m}= t_m/D_{R}$, i.e. the time spent by  
particles close to each other uninterruptedly. For our simulation, we choose again the distance $R_0$ to 
define proximity. 
Figure~\ref{fig:Expose_N_angle} displays the dependence of the scaled average exposure time $T_{m} Pe$  
on the dimensionless ratio $Pe^{\beta}/\Omega$ for particle density $\Phi=0.25$ . 
The relationship is studied for vision angles $\theta=\pi$, $\pi/2$, and $\pi/4$. The value of the exponent 
$\beta$, determined by good data collapse for different $Pe$ and $\Omega$,  depends on the vision angle, 
with $\beta=1$ for $\theta=\pi$, $\beta=2$ for $\theta=\pi/2$, and $\beta=-1/4$ for $\theta=\pi/4$. 
The qualitatively different scaling for $\theta \ge \pi/2$ and $\theta=\pi/4$ has the same origin as the
scaling of the average minimal distance $\langle d_1 \rangle$ in Fig.~\ref{fig:d1_angle}. 

Notably, the scaled exposure time becomes nearly independent of particle density or vision angle for 
$Pe^{\beta}/\Omega \gtrsim 1$, with
\begin{equation}
T_{m} Pe = A.
\end{equation}
The constant $A$ can be calculated in the ``ideal gas" limit of nearly straight particle trajectories, 
by considering the length of segments of straight lines intersecting a circle, with a homogeneous distribution 
of perpendicular distances from the circle center. This yields $A=A_{id}=\pi/2$, in reasonable agreement 
with the data in Fig.~\ref{fig:Expose_N_angle} for large $Pe^{\beta}/\Omega$. This indicates for $\theta=\pi$ and 
$\pi/2$ that the exposure time is nearly independent of steering and maneuverability at high particle 
velocities, and inversely proportional particle velocities $Pe$. 

Therefore, the interesting behavior, where particles can react to their environment by steering their motion, 
occurs for $Pe^{\beta}/\Omega \lesssim 1$.  For larger vision angle $\theta=\pi$ and $\pi/2$, the results
in Fig.~\ref{fig:Expose_N_angle} indicate that at higher maneuverability and lower activity, particles can 
steer well away from each other, so that the exposure time is very low. The exposure time is smaller for 
$\theta=\pi$ compared to $\theta=\pi/2$, which indicates better steering for particle avoidance. 
For the smaller vision angle $\theta=\pi/4$, the functional dependence of the exposure time reflects again
the qualitatively different behavior discussed above, with a maximal exposure time for 
$Pe^{-1/4}/\Omega \lesssim 3\times 10^{-2}$. 

The dependence of the average exposure time on activity, maneuverability, vision angle, and particle
density, reflects the motion and steering mechanisms discussed in the previous subsections. 
In particular, the prolonged exposure time for $\theta=\pi/4$ low $Pe^{-1/4}/\Omega$ can be 
attributed to collective motion in the form of bands within this regime. 
(for details see Sec.~\ref{Sec:Ped:Band}). 
Density has only a weak effect on exposure time, with somewhat longer exposure time at higher densities.
This is due to the definition of exposure time, where only particle pairs contribute which are within
the $R_0$ vision range.

\subsection{Mean-Square Displacement} 
\label{sec:msd}

\begin{figure}
    \centering
    \includegraphics[width=0.95\textwidth]{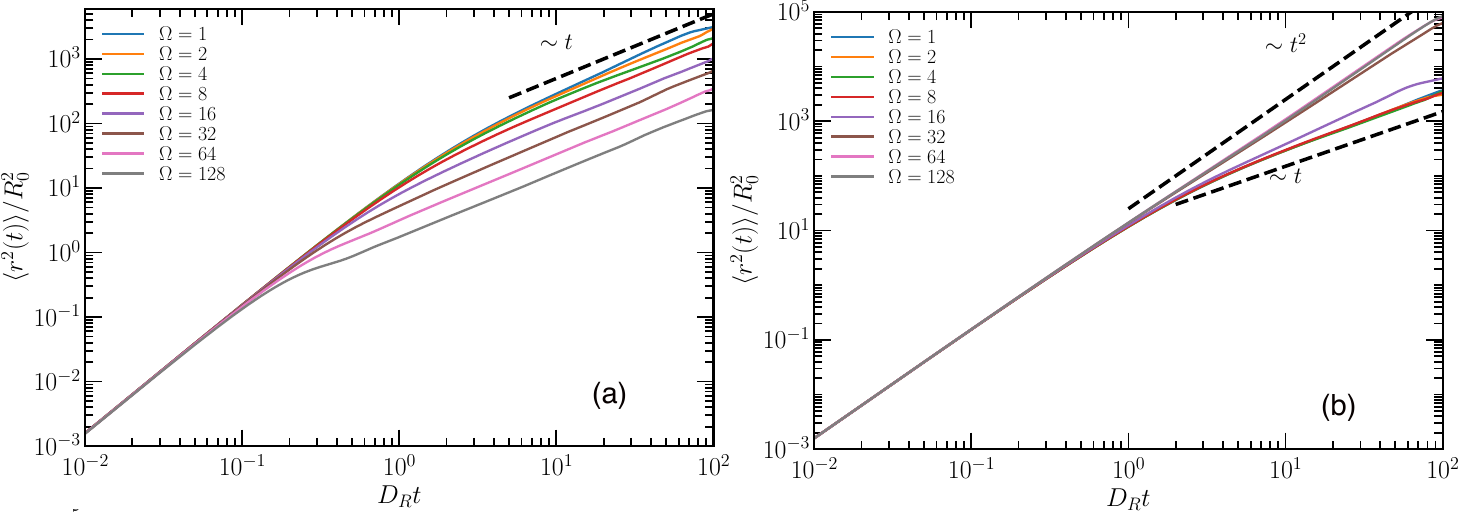}
    \caption{Mean squared displacement of particles at $\Phi=0.25$ and $Pe=4$ for various 
     maneuverabilities $\Omega$, as indicated, for vision angle (a) $\theta=\pi$ and (b) $\pi/4$.}
    \label{fig:msd_graph}
\end{figure}

The translational motion of the active Brownian particles is characterized by their mean-square 
displacement (MSD)
\begin{equation}\label{eq:MSD}
    \langle \boldsymbol r^{2}(t) \rangle= \frac{1}{N} \sum_{i=1}^N 
              \left\langle \left( \boldsymbol r_i(t+t_0)- \boldsymbol r_i(t_0) \right)^2 \right\rangle ,
\end{equation}
where the average is performed over the initial time $t_0$.
The theoretical calculations in two dimensions for active Brownian particles yield 
\cite{hows07, elgeti_2015_RPP, bechinger_2016_RMP}
\begin{equation}\label{eq:MSD_2}
	\langle \boldsymbol r^{2}(t) \rangle =4D_T t + \frac{2v_0^2}{D_R^2} \left( D_R t -1 + e^{-D_R t}\right) .
\end{equation} 
Figure \ref{fig:msd_graph}(a) displays the time dependence of the mean-square displacement for various 
maneuverabilies, with vision angle $\theta=\pi$ and fixed activity $Pe=4$. The particles exhibit short-time
ballistic and long-time diffusive behavior, where the effective translational diffusion coefficient decreases
with increasing maneuverability. The particles behave very similarly to free Active Brownian Particles (ABPs) 
for small maneuverability $\Omega=1$, while their diffusion is strongly reduced for large maneuverability 
$\Omega=128$, where particles are overly cautious in their movement and try to avoid the vicinity of their
neighbors. 

For vision angle $\pi/4$, the MSD displays two different power laws for long times, depending on the
maneuverability, see Figure \ref{fig:msd_graph}(b). 
For low maneuverability $\Omega \leq 8$, the MSD curves overlap, and the long-time MSD is diffusive with 
$MSD \sim t$. This behavior is typical of free ABPs. However, at higher maneuverability $\Omega \geq 32$, 
the long-time MSD is ballistic, with $MSD \sim t^2$. The latter case corresponds to the regime of 
$Pe^{-1/4}/\Omega \lesssim 3\times 10^{-2}$ in Figs.~\ref{fig:d1_angle} and \ref{fig:Expose_N_angle}, where 
band formation and collective motion of particles emerges. 

\begin{figure}
    \centering
    \includegraphics[width =.95\textwidth]{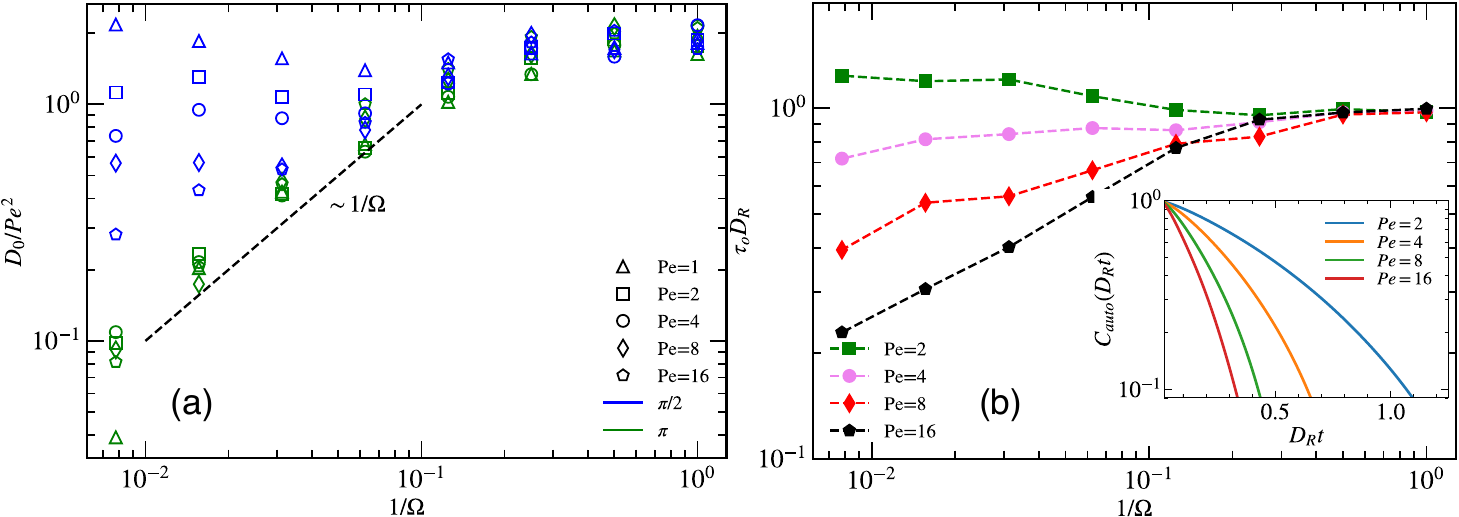}
    \caption{(a) Scaled effective long-time diffusion constant $D_{0}=D_{eff}/D_{R}$ extracted from the MSD at 
    various $Pe$, as indicated, for article density $\Phi=0.25$. (b) Relaxation time $\tau_{0}$ of particle motion direction as function of maneuverability $\Omega$, at 
    vision angle $\theta =\pi/2$ and particle density $\Phi=0.25$, for various $Pe$. 
    Inset: Auto-correlation function of the propulsion direction of individual particles at $\Omega=64$, vision 
    angle $\theta=\pi/2$, and particle density $\Phi=0.25$.}
    \label{fig:Effective_MSD_D}
\end{figure}

From the mean-square displacement (MSD) curve, we can derive the effective long-time diffusion constant 
for both vision angles $\pi$ and $\pi/2$, where we observe diffusive behavior. Results are presented in 
Fig.~\ref{fig:Effective_MSD_D}(a) as a function of maneuverability for particle density, $\Phi=0.25$, and various activities. 
To set the value of the diffusion coefficient $D_0= D_{eff}/D_{R}$ into perspective, 
we scale it with $Pe^2$, which corresponds to the behavior of free ABPs, compare Eq.~\eqref{eq:MSD_2}. 
Remarkably, data for different $Pe$ then collapse onto each other for small maneuverability $1/\Omega \geq 0.3$. 
The effect of particle density was not significant and qualitative similar behavior is obtained 
at higher density (see SI Fig.~S2). 

For systems with vision angle $\theta=\pi$, we observe a remarkable convergence of data points for different $Pe$. 
As $1/\Omega$ decreases,  the effective diffusion coefficient scales as $D_{0}/Pe^2 \sim 1/\Omega$, resulting
in the universal behavior $D_{0} \sim Pe^{2}/\Omega$, i.e., $D_{eff} \sim v_{0}^2/C_{0}$ independent of rotational 
diffusion. In contrast, data points for vision angle $\theta=\pi/2$
are very scattered for $1/\Omega \lesssim 0.3$, and no scaling behavior emerges.

\subsection{Temporal Auto-Correlation Function}

The temporal auto-correlation function of individual particles is given by 
\begin{equation}\label{eq:OCF}
     C_{auto}(t) = \frac{1}{N} \sum_{i=1}^N \left\langle  {\boldsymbol e}_i(t+t_0) \cdot {\boldsymbol e}_i(t_0) \right \rangle ,
\end{equation}
where $\boldsymbol{e}_i$ is the orientation of particle $i$, and $N$ is the total number of particles. 
The inset of Fig.~\ref{fig:Effective_MSD_D}(b) shows the auto-correlation function of particles in the strong-steering 
regime. The decay becomes faster with increasing P{\'e}clet number. Interestingly, this is in contrast to simple ABPs, 
where the relaxation time  is independent of $Pe$. This $Pe$-dependence is due to the closer encounters of particles 
at higher activity, which imply rapid changes of their orientation. Similar results were reported in the pedestrians 
experiment of Ref.~\cite{Echeverria_2021_SR}, where the direction of motion of faster-moving pedestrians also relax faster.

The relaxation time $\tau_{0}$ can be extracted from the initial exponential decay of the 
auto-correlation function,
\begin{equation}
     \sum_{i=1}^N \left\langle  {\boldsymbol e}_i(t+t_0) \cdot {\boldsymbol e}_i(t_0) \right \rangle = A \exp({-t/\tau_{0}}).
\end{equation}
Figure~\ref{fig:Effective_MSD_D}(b) illustrates the dependence of the relaxation time $\tau_{0}$ on $Pe$ and $\Omega$. 
For lower maneuverability, $1/\Omega \geq 0.1$, particles are in the ABP regime and hence the relaxation time is 
completely determined by rotational diffusion constant $D_{R}$. For higher maneuverabilities, where particles can 
steer effectively away from each other, the value of relaxation time is determined by both activity and maneuverability. 
Higher maneuverability results in stronger steering and consequently faster reorientation and relaxation and the 
propulsion direction.

\begin{figure}
    \centering
    \includegraphics[width=0.95\textwidth]{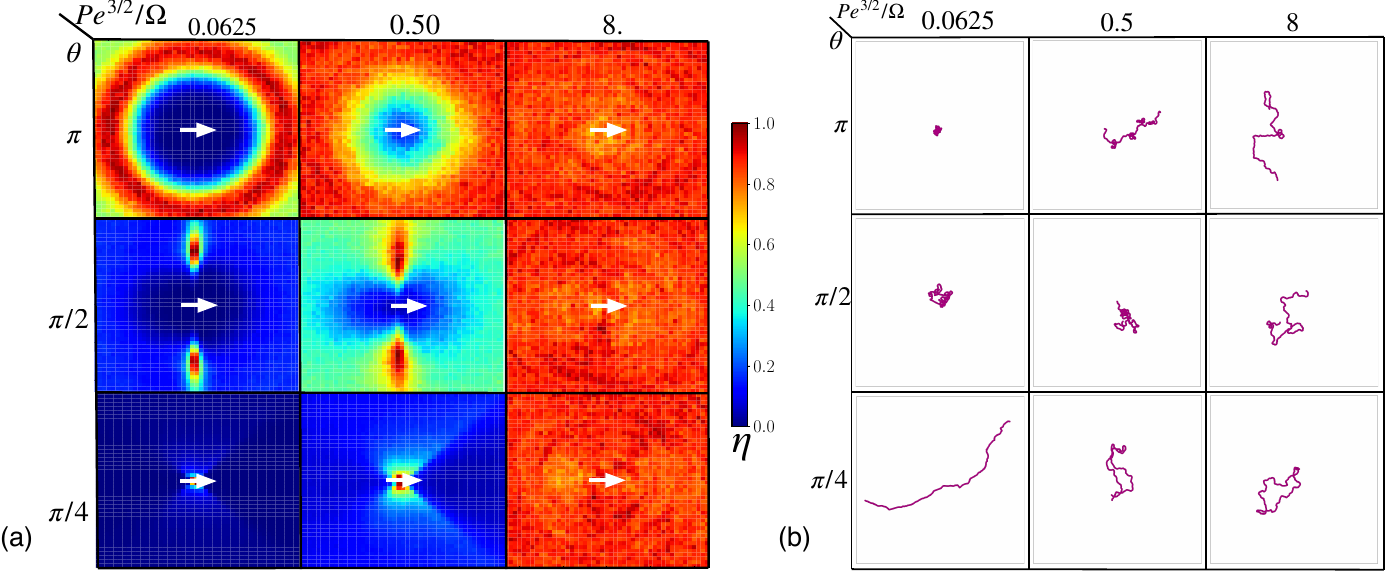}
    \caption{(a) Density distribution $\eta$ (normalized with the maximum value) of particles around a given particle 
    whose orientation is as indicated by
    the small white arrow to the right at the center, for various vision angles $\theta$ and activity-maneuverability
    ratios $Pe^{3/2}/\Omega$, as indicated.
    (b) Exemplary trajectories paths of active self-steering particles are shown for different vision angles $\theta$ and 
    P\'eclet-maneuverability ratios $Pe^{3/2}/\Omega$, as specified. All results are shown for $Pe=4$.}
    \label{fig:trajectories}
\end{figure}

\subsection{Local Particle Distributions and Trajectories}
\label{sec:density_distribution_trajectories}

To further characterize typical particle conformations and dynamics, we consider the density distribution in a 
particle-centered and oriented reference frame for various ratios $Pe^{3/2}/\Omega$ and vision angles $\theta$,
see Fig.~\ref{fig:trajectories}(a).
For vision angle $\theta=\pi$, the particle distribution is isotropic; for small $Pe^{3/2}/\Omega =0.0625$, there
is a pronounced density peak at the vision range $R_V$, indicative of high steering maneuverability, where 
particles are able to maintain a distinct separation from one another. This peak is smeared out and disappears with 
increasing $Pe^{3/2}/\Omega$. 
For vision angles $\theta=\pi/2$ and $\pi/4$, due to the asymmetry in the vision field, the density distribution 
also becomes highly asymmetric for small and moderate $Pe^{3/2}/\Omega$, with less number of particles in front 
and back, and more particles in the 
side-wise direction.

Figure~\ref{fig:trajectories}(b) displays corresponding representative trajectories. For vision angle $\theta=\pi$ 
and high maneuverability, with $Pe^{3/2}/\Omega < 0.1$, particles remain almost stationary,  just wiggling around 
their average location. As the vision angle decreases, and $Pe^{3/2}/\Omega$ increases, particle become more
mobile, and trajectories more persistent. 
Notable is the motions for $\theta=\pi/4$, where particles exhibit nearly straight and extended trajectories, 
which arises from the pronounced directional motion due to the formation of band-like structures 
(see Sec.~\ref{Sec:Ped:Band}).

\subsection{Band-like Structure at Narrow Vision Angles} 
\label{Sec:Ped:Band}

\begin{figure}
    \centering
    \includegraphics[width=.75\textwidth]{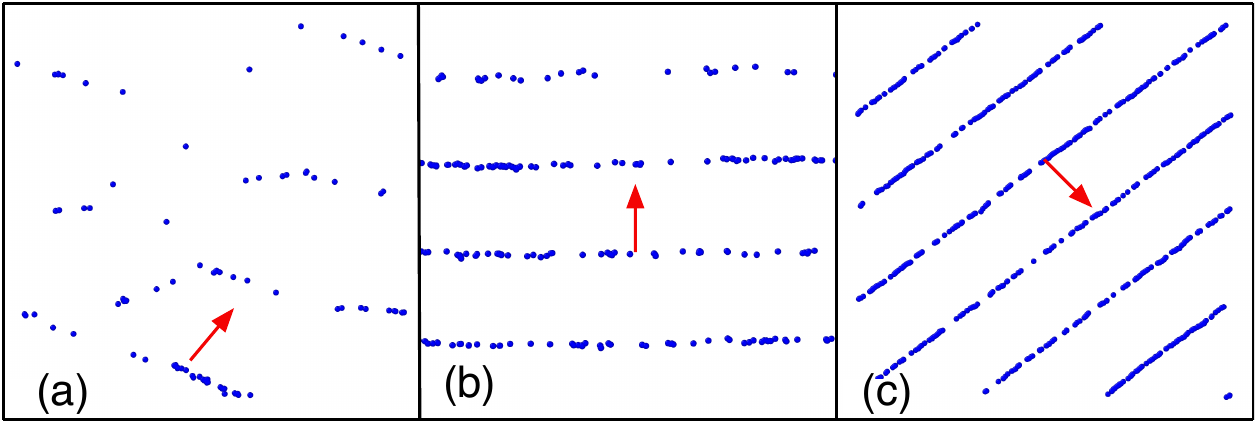}
    \caption{Snapshots showing band-like motion for vision angle $\pi/4$, for $Pe=4$, 
     $\Omega=16$, $R_{V}/R_{0}=4.0 $, and $Pe^{3/2}/\Omega=0.125$, for particle's density 
     (a) $\Phi=0.25$, (b) $\Phi=0.625$, and (c) $\Phi=2.5$. 
     The red arrow indicates the propagation direction of the particle. See also movies M3, M4.}
    \label{fig:band_strc}
\end{figure}

As noted above, band-like aggregates and motion patterns appear for low activity-maneuverability 
ratio $Pe^{3/2}/\Omega=0.125$ and narrow vision angle $\theta= \pi/4$, reminiscent of the bands 
in the Vicsek model near the transition from the polarized to the disordered phase 
\cite{vicsek_1995_PRL, vicsek_2012_PhysRep, gregoire_2004_PRL}. 
However, these bands are very thin compared to the bands in the Vicsek model. The restricted vision  
implies that the particles can only react to and interact with other particles in front of them, 
but are not aware of or responsive to particles on their sides in perpendicular directions; 
thus, the particles can come very close to each other, with small distances to the nearest neighbors 
$\langle d_1 \rangle$ (see Sec.~\ref{sec:ped:Distance_NND1}) and a large exposure times $T_{m}$ 
(see Sec.~\ref{sec:Exposure_T}). 
Figure \ref{fig:band_strc} shows typical snapshots of band-like structures at different particle 
densities. When the particle density is low, $\Phi = 0.25$, the band-like structures are not 
very prominent, because the particles have more available space to move around, allowing for more freedom of 
motion. However, as the density increases, the 
available space per particle decreases, and the band-like structures become much more distinct, 
even forming a one-dimensionally ordered stripe phase at $\Phi=2.5$. 

\subsubsection{Polarization} 
\label{sec:polarization}

\begin{figure}
    \centering
    \includegraphics[width =.95\textwidth]{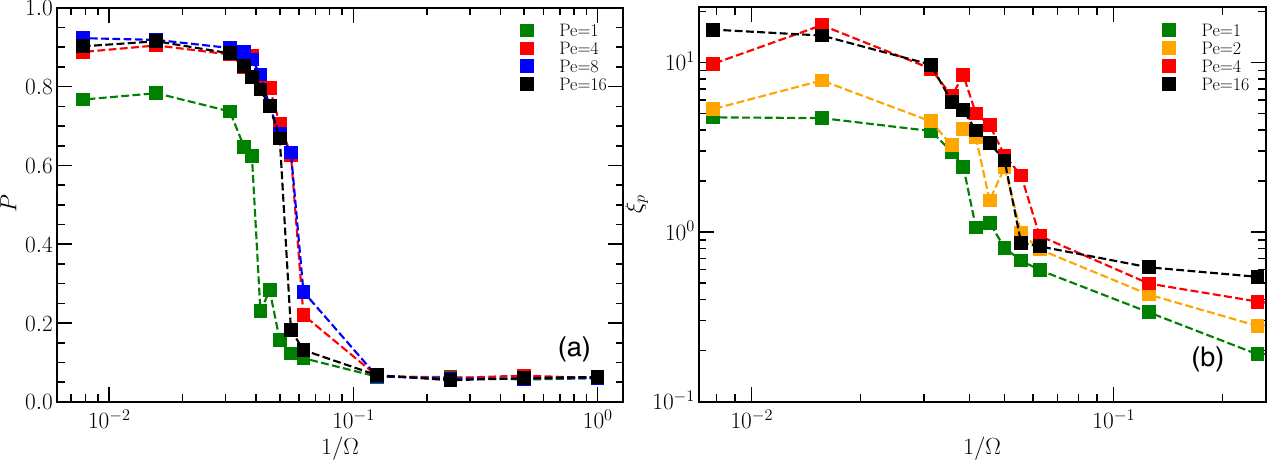}
  \caption{  (a) Polarization $P$ as a function of maneuverability $\Omega $ at 
     vision angle $\pi/4$, density $\Phi=0.625$ at various $Pe$, as indicated. The sharp drop of 
     $P$ at $1/\Omega \simeq 0.05$ indicates a transition from the uniformly distributed, randomly 
     moving ABPs at low $\Omega$ to collective motion in the form of bands at high $\Omega$. (b) Persistence length $\xi_{p}$ extracted from the spatial correlation function for 
    $\Phi=0.625$, and vision angle $\pi/4$, for various indicated $Pe$ as a function of maneuverability 
    $\Omega$, showing the transition from randomly moving ABPs having low persistence motion to highly 
    persistence collective motion. }
    \label{fig:ped:Polarization}
\end{figure}

We characterize the transition from the state of disordered motion to band formation by
the global polarization order parameter \cite{aitor_2018_soft_matter,negi2023collective}
\begin{equation} \label{eq:polarization}
	P=  \left\langle \frac{1}{N} \left| \sum_i \mathbf{e}_i \right| \right\rangle ,
\end{equation}
where ${\boldsymbol e}_i$ is orientation of particle $i$ and the average is performed over time.
Figure~\ref{fig:ped:Polarization}(a) illustrates the polarization $P$ as a function of 
maneuverability $\Omega$, at particle density of $\Phi=0.625$. At low maneuverability, 
i.e. for $1/\Omega \geq 1/8 $, particles display random orientations, resulting in polarization 
$P \approx 0$. 
However, as maneuverability increases, a transition occurs at $1/ \Omega \simeq 1/16$,
where particles align their orientations and a banded state with large global 
polarization emerges. 
As $\Omega$ increases further, the polarization nearly reaches unity, in particular 
for $1/\Omega \leq 1/32 $, and larger $Pe$.

\subsubsection{Spatial Correlations and Persistence Length}
\label{sec:ped:spatial_}

Another interesting quantity to characterize the banded state is
the spatial correlation function \cite{doliwa_2000_PRE, wysocki_2014_EPL},
\begin{equation}
\label{eq:spatial}
    C_{e}(\boldsymbol r) =\left\langle \frac{\displaystyle  
            \sum_{i,j\ne i}  \boldsymbol e_i  \cdot \boldsymbol e_j \delta(\boldsymbol r-(\boldsymbol r_i-\boldsymbol r_j)) }
    {\displaystyle   \sum_{i,j\ne i}  \delta(\boldsymbol r-(\boldsymbol r_i-\boldsymbol r_j)) }\right\rangle ,
\end{equation}
where $\boldsymbol e_i$ and $\boldsymbol e_j$ represent the orientation vectors for particle $i$ and particle $j$, 
respectively. The spatial correlation function can be used to extract the information about the 
persistence length $\xi_p$ as
\begin{equation}
    \left\langle  \sum_{i,j\ne i}  \boldsymbol e_i  \cdot \boldsymbol e_j \delta(\boldsymbol r-(\boldsymbol r_i-\boldsymbol r_j))  
     \right\rangle= B \exp ({-|\boldsymbol{r}_{i}-\boldsymbol{r}_j|/ \xi_p}) .
\end{equation}
The persistence length of the band-like structures is shown in 
Fig.~\ref{fig:ped:Polarization}(b) for different activities $Pe$ and 
maneuverabilities $\Omega$, at $\Phi=0.625$. A transition from a high-persistence-length phase 
at $1/ \Omega \leq 1/32$ to a low-persistence-length phase at $1/ \Omega \geq 1/16$ is evident, 
similar to the behavior observed for global polarization \ref{sec:polarization}. The dependence 
on the P\'eclet number ($Pe$) appears to be relatively weak, with $\xi_p$ decreasing with 
decreasing $Pe$.

\section*{Discussion}
We have analyzed the behavior of active, self-steering particles with visual perception in 
semi-dilute crowds, where each particle self-steers to avoid regions of high neighbor density in 
their vision cone. We focus on a minimal model, where particles move with constant velocity
and have only instantaneous spatial information of their neighbors.

The dependence of the probability density function (PDF) of the 
minimal distance $d_1$ to neighboring  particles, the fraction of particles in the close vicinity of
neighbors, as well as the expose time on particle P{\'e}clet number $Pe$, maneuverability $\Omega$, 
vision angle $\theta$, and density $\Phi$ is investigated. We find that the PDF $P(d_1)$ displays a peak, 
which shifts toward lower value with increasing particle density. Furthermore, PDFs for constant 
$Pe^{3/2}/\Omega$ ratio display universal scaling behavior, which indicates that stronger maneuverability 
is required at high activities to avoid close contact; similar features were observed in previous studies 
of iABP \cite{goh_2022_NJP}, and iAOUP (intelligent Active Ornstein Uhlenbeck particles) pursuit dynamics 
\cite{gassner_2023_epl}, where self-steering is {\em toward} regions of high local particle density 
in the vision cone.

We also examined the impact of the vision angle $\theta$, which reveals that particles with wider 
fields of view (larger $\theta$) are better equipped to detect potential collisions and steer away 
from potential collisions earlier, resulting in a larger minimum distance $d_1$. Conversely, 
particles with narrower fields of view have shorter minimum distances. For narrow vision cone (vision angle 
$\theta=\pi/4$), we observe the formation of band-like structures and collective motion for high 
maneuverability strength, somewhat reminiscent of the bands in the Vicsek model.  

For the duration that particles spent in close proximity to other particles, known as exposure time 
$T_{m}$, we find a correlation between high levels of particle activity and short exposure time. For large vision angles $\pi$ and $\pi/2$, the scaled exposure $T_m Pe$ show a consistent universal behavior as function the activity-maneuverability ratio $Pe^{\beta}/\Omega$, with $\beta=1$ and $\beta=2$ respectively. For small vision angle $\pi/4$, exposure time is very high at high maneuverability and is characterized by a negative exponent $\beta=-1/4$, due to the formation of band-like structures. The motion of agents within these bands is highly persistent, as indicated by their trajectories 
and persistence lengths. Furthermore, these bands are associated with a highly polarized state, characterized by a polarization order parameter $P \approx 1$.

The results of our model system can be compared -- to some extent -- to those of recent experiments
of walking pedestrians confined in a room, with the goal to maintain a large ``safety'' distance to other 
pedestrians \cite{Echeverria_2021_SR}. Several of our results are in good qualitative agreement with
the experimental observations.
As the pedestrian density increases, interactions become more frequent, leading to smaller distances 
between them.  Additionally, more briskly walking pedestrians exhibit reduced minimum distances, as higher 
activity requires pedestrians to approach others more closely before steering becomes effective to avoid 
collisions \cite{Echeverria_2021_SR}.  
The experimental results on fast-moving pedestrians also reveal similar features of exposure time as in 
our simulations, with an exposure time that is inversely proportional to walking speed 
\cite{Echeverria_2021_SR}. We also observe faster relaxation of orientation direction for higher particles 
activity, in good agreement with experiments on pedestrians \cite{Echeverria_2021_SR}.

We want to emphasize that our model has of course several limitations in describing the behavior of
real pedestrians. One limitation is the idealization of constant speed, while pedestrians can adapt their
speed. Another is that we consider instantaneous spatial information only, while pedestrians are able
to judge the motion direction and speed of their neighbors, extrapolate to future collision points,
and adjust their motion accordingly. However, such extrapolations become increasingly difficult as
the number of particles in the vision cone increases.

Overall, our study sheds light on the complex interplay between particle behavior, activity levels, vision 
angles, and other parameters. For large vision angles, like $\theta=\pi$ and $\pi/2$, the results of our
model can qualitatively match some behavior of pedestrian crowds like mean exposure time and the probability 
distribution function of distance to the nearest neighbor. 
It would certainly be interesting to study the distance distributions of in other animal swarms more quantitatively,
where flocks of birds and swarms of insects look like promising candidates.

\section*{Acknowledgements}

We would like to thank Mohcine Chraibi (Forschungszentrum J\"ulich) for introducing 
us to the distancing problem in human crowds. 

\section*{Author contributions statement}

G.G. designed the project, R.S.N.  wrote the code and performed the simulations. R.S.N. and P.I. analyzed the results. All authors contributed to the discussions. R.S.N. and G.G. wrote the manuscript. All authors contributed to reviewing the manuscript.


\end{document}